\documentclass{article}
\usepackage{arxiv}
\pdfoutput=1

\usepackage[utf8]{inputenc} 
\usepackage[T1]{fontenc}    
\usepackage{url}            
\usepackage{amsfonts}       
\usepackage{nicefrac}       
\usepackage{microtype}      
\usepackage{textcomp}
\usepackage{amsmath}
\usepackage{placeins}
\usepackage{graphicx}
\usepackage{subcaption}
\usepackage{mwe}

\title{Geometry of the EOS\textregistered\ Radiographic Scanner}

\author{
  Benjamin N.~Groisser \\
  Department of Mechanical Engineering\\
  Technion-Israel Institute of Technology\\
  Haifa, Israel \\
  \texttt{bgroisser@campus.technion.ac.il}
}

\begin{document}
\maketitle

\begin{abstract}
The EOS\textregistered\ scanner is a radiographic system that captures PA and lateral images in standing posture. The system is widely used in diagnosis and assessment of scoliosis, as it provides a low-dose alternative to traditional X-ray and can capture full-body images. Furthermore, spacial calibration between the two imaging views is implemented in hardware, facilitating 3D reconstruction of imaging targets. In this paper, a brief description of the system is followed by an explanation of the geometric relationship between 3D space and radiographic image space. 
\end{abstract}

\section{Introduction}
The EOS\textregistered\ scanner is a unique radiographic system with several key features that have made it an key tool for diagnosis and assessment of scoliosis, in addition to use in assessment of e.g. knee arthritis. 

The physics of the multiwire chamber used to amplify the primary X-ray emissions are beyond the scope of this document; likewise, treatment of the clinical use of the system can be found elsewhere\cite{Illes2012}. Rather, what follows is a description of the relevant radiographic components and their relative positioning, which will then be used to derive simple formulae for projection and reconstruction between 3D space and image space.

\section{Hardware}
\label{sec:hardware}

\subsection{Slot Scanners}
Radiographic slot scanners are an alternative to standard digital radiography. While both technologies will produce a two dimensional image, there are significant differences in construction that determine the properties of the output images.

\begin{figure*}
    \centering
    \begin{subfigure}[b]{0.6\textwidth}
        \centering
        \includegraphics[height=8cm]{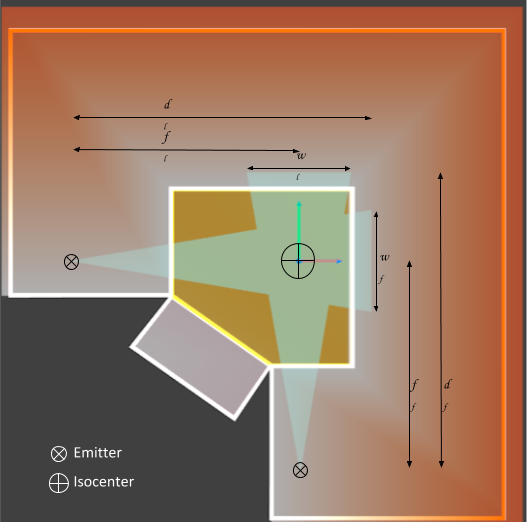}
        \caption[]%
        {{\small Orthographic overhead view.}}    
    \end{subfigure}
    \begin{subfigure}[b]{0.3\textwidth}   
        \centering 
        \includegraphics[height=8cm]{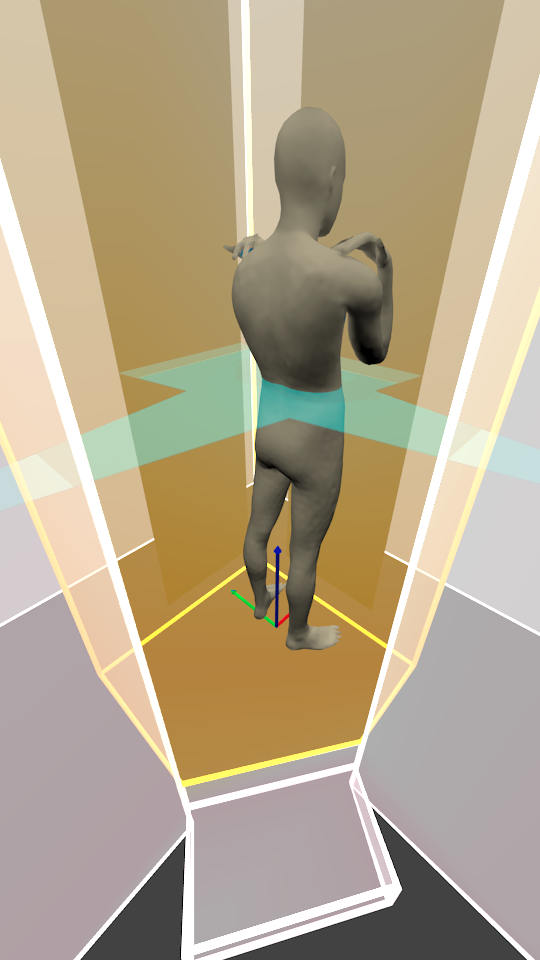}
        \caption[]%
        {{\small Perspective view}}    
    \end{subfigure}
    \caption[Synthetic projections using clinical CT data]
    {\small Renderings of EOS radiographic environment. Global coordinate axes are displayed at the origin: X (red) towards the frontal detector, Y (green) towards the lateral detector, and Z (blue) vertically upright. For details on each symbol, consult Appendix \ref{sec:appendix}.} 
    \label{fig:render}
\end{figure*}

In the EOS\textregistered\ scanner, X-ray emitters and detectors are mounted on an enclosed gantry. The emission beam profile is a highly collimated fan restricted to the axial plane; emitter and detector move in unison during the scan to traverse the field of view. Axial scan lines are then assembled to form a two dimensional image. This axial beam profile and matching detector has the effect of filtering out deflected radiation, helping to reduce imaging noise.

\subsection{Geometry}

Two X-ray emitters (with paired detectors) are mounted orthogonally. These elements are all linked to move in unison, such that frontal and lateral images are collected concurrently. The relative positions of the emitters and detectors are fixed and calibrated at installation.

Note that each row of the image is perpendicular to the direction of travel of the gantry, and that each row on the frontal image corresponds to the same row in the lateral image. This implies that, for any given point on one radiographic image, the epipolar line is exactly the corresponding row on the other image.

Also note that, due to the fan pattern of the emission beam, the effective image width at the isocenter is smaller than the physical detector width. The specific model used in this discussion generates square pixels; as such, images are true-scale \textit{on the plane of the isocenter}. Structures located closer or farther from the X-ray tube will experience magnification and parallax distortion.

\section{Projections and Reconstructions}

As a matter of convention, in the following sections image rows will be numbered running from $0$ as the top row to $R$ as the bottom row. Likewise, columns run from $0$ on the left edge to $C$ on the right. A list of all the relevant parameters, along with representative values, can be found in Appendix \ref{sec:appendix}. Note that, although the physical detectors for frontal and lateral images are typically equal size, the difference in distance from the isocenter results in a larger effective image width for the frontal image. 

\subsection{Projection}

Mapping points from 3D world space to radiographic image space can be achieved with a modified pinhole camera model. The pixel row in the frontal image $v_f$ is simply the vertical height of the point $P_z$ scaled by the pixel pitch $\lambda_z$:

\begin{equation}
    v_f = \frac{z_0 - P_z}{\lambda_z}
\end{equation}

where $z_0$ is the initial height of the emitter. To find the projected location in homogeneous coordinates $u'$, transform the 3D point $P$ in global coordinates into camera coordinates, then project into 2D with scaling factor $\omega$. 

\begin{equation}
    \begin{bmatrix} u_f' \\ \omega \end{bmatrix} = \\
        \begin{bmatrix} f_f & 0 & 0 & 0 \\ 0 & 0 & 1 & 0 \end{bmatrix} 
        \begin{bmatrix} 0 & 1 & 0 & 0 \\ 
                        0 & 0 & 1 & -P_z \\ 
                        1 & 0 & 0 & f_f \\
                        0 & 0 & 0 & 1 \end{bmatrix} 
        \begin{bmatrix} P_x \\ P_y \\ P_z \\ 1 \end{bmatrix} =
        \begin{bmatrix} P_y f_f \\ f_f + P_x \end{bmatrix}
\end{equation}

Converting from homogeneous coordinates to column index is performed by scaling by horizontal pitch, then offset to begin indexing from the image edge:

\begin{equation}
\label{eq:latproj}
    u_f = \frac{C_f}{2} + \frac{P_y f_f}{\lambda_f (f_f + P_x)}
\end{equation}

The same process can be used to find the pixel coordinates for the lateral image:

\begin{equation}
\renewcommand*{\arraystretch}{2}
\begin{bmatrix} u_l \\ v_l \end{bmatrix} = 
    \begin{bmatrix} \frac{C_l}{2} - \frac{P_x f_l}{\lambda_l (f_l + P_y)} \\
                    (z_0 - P_z)/\lambda_z\end{bmatrix}
\end{equation}

Note that the sign of the offset (e.g. second term in Equation \ref{eq:latproj}) will determine the orientation of the projected image. The convention used here is radiographic standard: image is viewed from ``behind'' the imaging screen (e.g. in a posterior-anterior scan the patient left will appear on image right).

\subsection{Reconstruction}
\label{sec:reconstruction}

Radiographic reconstruction can be seen as the inverse of projection; points on lateral and frontal radiographs are reconstructed to find the corresponding point in 3D space. This is performed by back-projecting the points from the imaging screen to the X-ray source and finding the intersection between frontal and lateral projection lines. This can be formulated as solving a system of equations. First, convert pixel coordinates into 3D location:

\begin{equation}
\begin{matrix}
    x_f =& 0                                            &,\ \ \ \  &  x_l =& \lambda_l \left( \frac{C_l}{2} - u_l \right) \\
    y_f =& \lambda_f \left( u_f - \frac{C_f}{2}\right) &,\ \ \ \  &  y_l =& 0 
\end{matrix}
\end{equation}

Then solve the simultaneous equations

\begin{align}
 \renewcommand*{\arraystretch}{1.8}
    P_y &= \frac{y_f}{f_f} \left( P_x - f_f \right) \nonumber\\ 
    P_x &= \frac{x_f}{f_l} \left( P_y - f_l \right) \nonumber\\
\Rightarrow   \begin{bmatrix} -y_f & f_f \\
                    f_l & -x_l \end{bmatrix}
    \begin{bmatrix} P_x \\ P_y \end{bmatrix}
    &= \begin{bmatrix} f_f y_f \\ f_l x_l \end{bmatrix} 
\end{align}

Solving for $P_x, P_y$ can be easily performed e.g. with matrix inversion. Solving for $P_z$ is straightforward. If, due to rounding or labeling errors, the row index for the lateral and frontal images are not equal, a simple average can be used:

\begin{equation}
    P_z = z_0 - \lambda_z \frac{v_l + v_f}{2}
\end{equation}

\section{Examples}

\subsection{Synthetic Radiographs from CT}

\begin{figure*}
    \centering
    \begin{subfigure}[b]{0.22\textwidth}
        \centering
        \includegraphics[height=9cm]{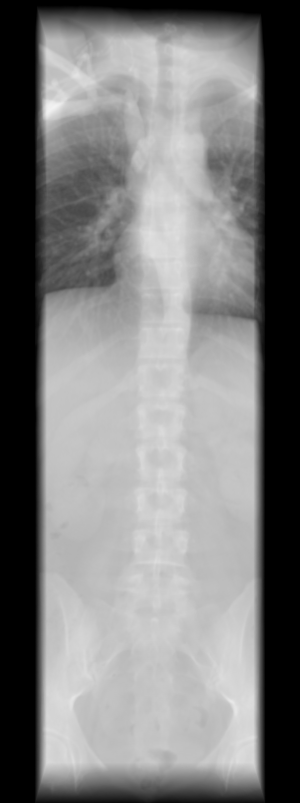}
        \caption[]%
        {{\small DR PA}}    
    \end{subfigure}
    \begin{subfigure}[b]{0.22\textwidth}   
        \centering 
        \includegraphics[height=9cm]{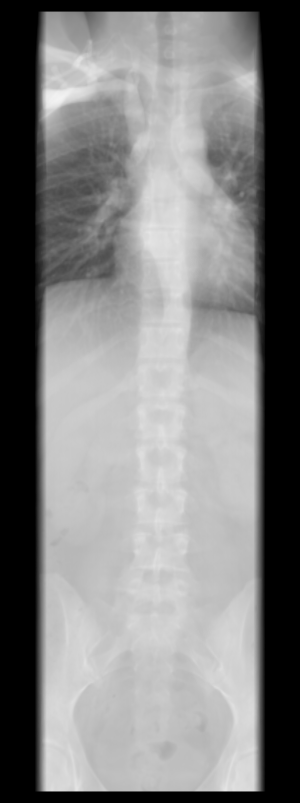}
        \caption[]%
        {{\small EOS PA}}    
    \end{subfigure}
    \begin{subfigure}[b]{0.22\textwidth}  
        \centering 
        \includegraphics[height=9cm]{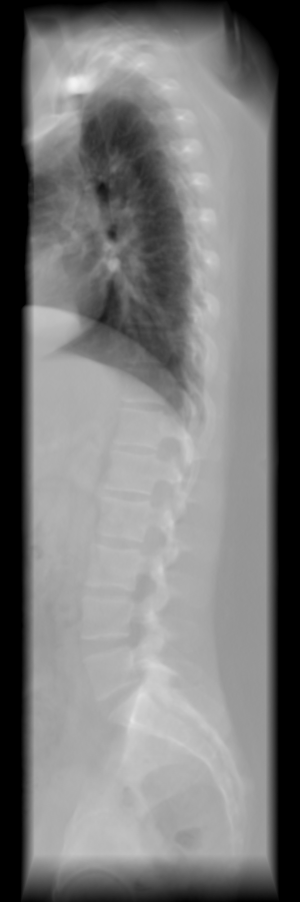}
        \caption[]%
        {{\small DR LAT}}    
    \end{subfigure}
    \begin{subfigure}[b]{0.22\textwidth}   
        \centering 
        \includegraphics[height=9cm]{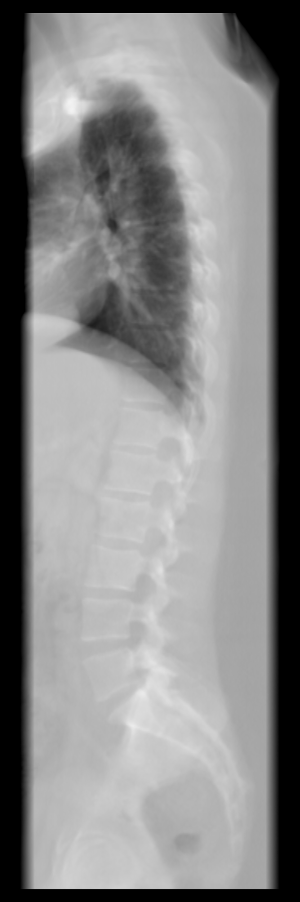}
        \caption[]%
        {{\small EOS LAT}}    
    \end{subfigure}
    \caption[Synthetic projections using clinical CT data]
    {\small Synthetic projections using clinical CT data. Top row shows Digital Radiography while bottom row demonstrates EOS geometry. In the case of DR the X-ray source (i.e. pinhole) is positioned in the center of the spine. Emitter distance to patient and detector are maintained between the two trials.} 
    \label{fig:DR_vs_EOS}
\end{figure*}

In Figure \ref{fig:DR_vs_EOS}, synthetic radiographs are generated from a clinical CT volume. Data is provided by []. The same volume is projected using two geometries: first, using a standard pinhole model to simulate standard Digital Radiography, and second, using the geometry described above. For each pixel in the image, the emission line from X-ray tube to detector is constructed. The segment that passes through the CT volume is interpolated and integrated to compute the pixel value\footnote{It is possible to simulate different X-ray wavelengths by adjusting the integral. For example, adding an aluminum filter might be modeled by applying a high-pass filter to the CT data, to selectively amplify the contribution of dense material.}. 

\subsection{Reconstructing Corresponding Points}

\begin{figure*}
    \centering
    \begin{subfigure}[b]{0.19\textwidth}
        \centering
        \includegraphics[height=6.5cm]{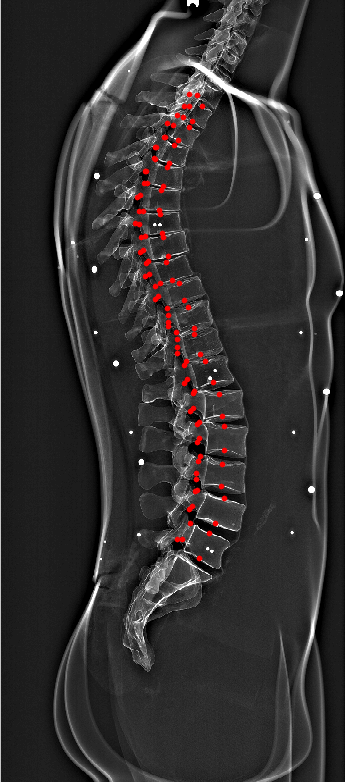}
        \caption[]%
        {{\small PA landmarks}}    
    \end{subfigure}
    \begin{subfigure}[b]{0.19\textwidth}   
        \centering 
        \includegraphics[height=6.5cm]{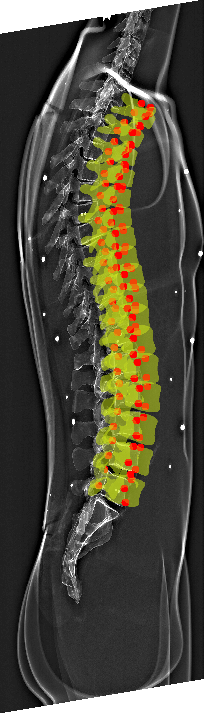}
        \caption[]%
        {{\small LAT + model}}    
    \end{subfigure}
    \begin{subfigure}[b]{0.19\textwidth}  
        \centering 
        \includegraphics[height=6.5cm]{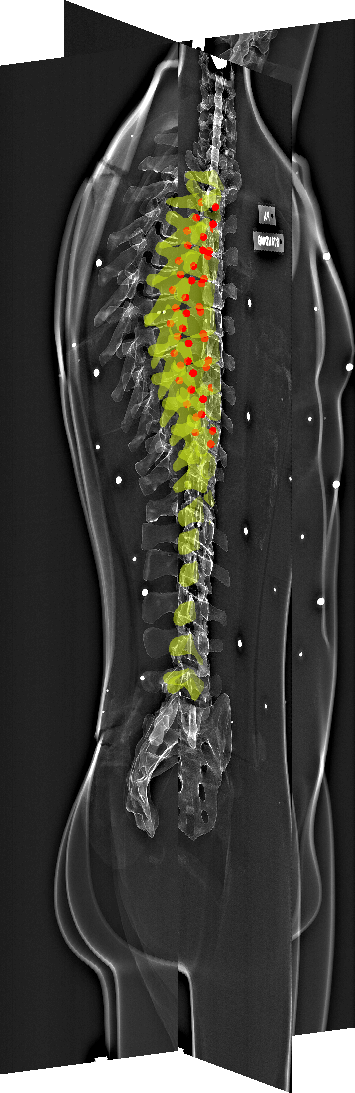}
        \caption[]%
        {{\small PA,LAT + model}}    
    \end{subfigure}
    \begin{subfigure}[b]{0.19\textwidth}  
        \centering 
        \includegraphics[height=6.5cm]{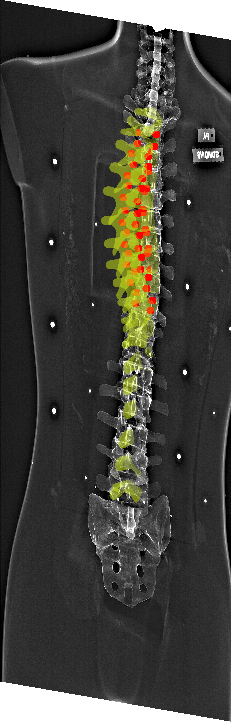}
        \caption[]%
        {{\small PA + model}}    
    \end{subfigure}
    \begin{subfigure}[b]{0.19\textwidth}   
        \centering 
        \includegraphics[height=6.5cm]{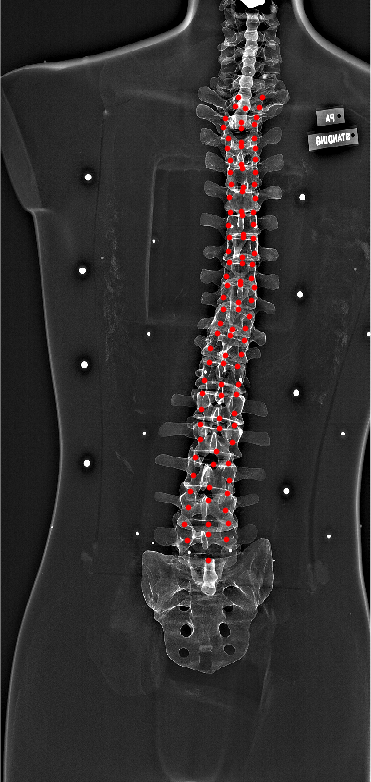}
        \caption[]%
        {{\small PA landmarks}}    
    \end{subfigure}
    \caption[Reconstructions of stereo-corresponding points.]
    {\small Reconstructions of stereo-corresponding points. A spine model is also fitted for visualization. Note that these images are flipped relative to radiographic convention, for better alignment with 3D space.} 
    \label{fig:reconstructions}
\end{figure*}

An important use of stereo-radiographic systems is reconstructing 3D structures from planar images. One commonly used method is to find corresponding points in both images and then recover the 3D position using the process described in Section \ref{sec:reconstruction}. Typically, the difficult part of this process is to locate corresponding points in both images; this is often a challenging task requiring expert training. In spinal vertebrae, six points have been identified as readily identifiable in frontal and lateral projections\cite{Andre1994}. In Figure \ref{fig:reconstructions} these points have been manually labelled for thoracic and lumbar vertebrae of a phantom spine model. These landmarks are then reconstructed, and a vertebral mesh model is registered with six degrees of freedom (translation, rotation) to fit each set of landmarks. All vertebral mesh models are rendered together in 3D, with radiographic images positioned at the isocenter.

\section{Conclusions}
The EOS\textregistered\ scanner is a hardware calibrated stereo radiographic system. The known geometry of the system facilitates projections from world coordinates to image coordinates, or reconstructions of corresponding landmarks from image space to world space. This is particularly useful for assessment of 3D anatomical structures in standing posture, such as the spine or lower extremities.

\newpage

\appendix
\section{Parameters}
\label{sec:appendix}
As a representative example, the following parameters describe the EOS\textregistered \ system installed in the Pediatric Radiology department at the Hospital for Special Surgery (HSS)\footnote{Hospital for Special Surgery - 535 E 70th St, New York City, New York 10021}.

\begin{enumerate}\interlinepenalty10000 
    \item $d_f$ : 1300mm - distance from frontal emitter to detector
    \item $f_f$ : 987mm - distance from frontal emitter to isocenter
    \item $d_l$ : 1300mm - distance from lateral emitter to detector
    \item $f_l$ : 918mm - distance from lateral emitter to isocenter
    \item $w_f$ : 450mm - width of frontal detector
    \item $w_l$ : 450mm - width of lateral detector
    \item $\lambda_x$ : 0.179363mm - horizontal pixel pitch for frontal image
    \item $\lambda_y$ : 0.179363mm - horizontal pixel pitch for lateral image
    \item $\lambda_z$ : 0.179363mm - vertical pixel pitch (shared by construction between frontal and lateral)
    \item $R$ : varies - number of rows in each images (typically 7,000-10,000 for a full-body scan)
    \item $C_f$ : 1895 - highest column index in frontal image (note there are $C_f+1$ total columns)
    \item $C_l$ : 1763 - highest column index in lateral image (note there are $C_l+1$ total columns)
\end{enumerate}

\bibliographystyle{unsrt}  


\end{document}